%% file: RPTU_paper.tex
\documentclass[conference]{IEEEtran}
\IEEEoverridecommandlockouts

\usepackage{cite}
\usepackage{amsmath,amssymb,amsfonts}
\usepackage{comment}
\usepackage{algorithmic}
\usepackage{graphicx}
\usepackage{textcomp}
\usepackage{booktabs,caption}
\usepackage[flushleft]{threeparttable}
\usepackage{xcolor}
\def\BibTeX{{\rm B\kern-.05em{\sc i\kern-.025em b}\kern-.08em
    T\kern-.1667em\lower.7ex\hbox{E}\kern-.125emX}}
    
\usepackage{algorithm}
\usepackage{array}
\usepackage[caption=false,font=normalsize,labelfont=sf,textfont=sf]{subfig}

\usepackage{stfloats}
\usepackage{url}
\usepackage{verbatim}
\usepackage{acro}

\include{acronyms}

\begin{document}

\title{Bridging Simulation and Measurements through Ray-Launching Analysis: A Study in a Complex Urban Scenario Environment


\thanks{This work has been funded by the Federal Ministry of Education and Research of Germany (BMBF). This is a preprint version, the full paper has been accepted by the 13th edition of the International Conference on Electromagnetics in Advanced Applications (ICEAA 2024), Lisbon, Portugal, September 2024. Please cite as: X. Long, A. Fellan, M. Pauli, H. D. Schotten, and T. Zwick, “Bridging Simulation and Measurements through Ray-Launching Analysis: A Study in a Complex Urban Scenario Environment,” 2024 International Conference on Electromagnetics in Advanced Applications (ICEAA 2024)}}

\author{\IEEEauthorblockN{Xueyun Long\textsuperscript{$*$}, Amina Fellan\textsuperscript{$\diamond$}, Mario Pauli\textsuperscript{$*$}, Hans D. Schotten\textsuperscript{$\diamond$}, Thomas Zwick\textsuperscript{$*$}}
\IEEEauthorblockA{\textsuperscript{$*$}\textit{Institute of Radio Frequency Engineering and Electronics (IHE)} \\
\textit{Karlsruhe Institute of Technology (KIT)}, Karlsruhe, Germany \\ xueyun.long@kit.edu}
\IEEEauthorblockA{\textsuperscript{$\diamond$}\textit{Institute of Wireless Communication and Navigation} \\
\textit{Rheinland-Pf\"{a}lzische Technische Universit\"{a}t Kaiserslautern-Landau (RPTU)}, Kaiserslautern, Germany \\ fellan@eit.uni-kl.de}
}

\maketitle

\begin{abstract}
With the rapid increase in mobile subscribers, there is a drive towards achieving higher data rates, prompting the use of higher frequencies in future wireless communication technologies. Wave propagation channel modeling for these frequencies must be considered in conjunction with measurement results. This paper presents a ray-launching (RL)-based simulation in a complex urban scenario characterized by an undulating terrain with a high density of trees. The simulation results tend to closely match the reported measurements when more details are considered. This underscores the benefits of using the RL method, which provides detailed space-time and angle-delay results.
\end{abstract}


\section{Introduction} \label{intro}
Wireless communication technologies are permeating through our lives. The proliferating number of mobile service subscribers triggered a demand for higher bandwidths and larger capacities that ensures good \ac{QoS} and \ac{QoE} for the end users. As a result, mobile wireless technologies addressed this demand by pursuing higher frequencies within the radio spectrum. In \ac{5G}, for instance, new operating frequencies were defined in the sub 6 GHz and the 24.25 GHz to 71.0 GHz \ac{mmWave} frequency ranges, also known as \ac{FR1} and \ac{FR2}. Consequently, a better understanding of radio waves and their channel propagation models at these frequencies is essential to optimize network planning and deployment of \acp{BS}. It is also pertinent for comprehending the nature of \ac{EMF} exposure in a given environment, a topic that has stirred public concerns with the deployment of \ac{5G} technologies in recent years. 

There are mainly two approaches to gain insights of the channel propagation. One of the simplest and fastest methods is to directly use empirical channel models, which are based on statistical models summarized from a large number of measurement campaigns in different environments, e.g., urban, suburban, or rural areas, and thus they prove useful as efficient channel approximations. However, they tend to be more accurate in environments similar to those from which the models were derived. Moreover, they are not capable of providing precise channel parameters information such as delay and angular spreads. 
 
The second approach to examine wave propagation deterministically is using \ac{GO}-based methods such as \ac{RT} and \ac{RL}. These simulations can numerically solve Maxwell equations to predict all the possible paths between a transmitter and a receiver. Based on 3D environments, field strength, time delay, \ac{AoD} and \ac{AoA} for each paths will be precisely derived involving different wave propagation phenomena. Therefore they have widespread applications. The cost, however; is a significant computational effort and long computation times, such as IHE-\ac{RT} in \cite{fugen2006capability} which utilizes image method. For instance, to identify reflection paths, an image of the receiver point is generated relative to all planes in the scenario. Connecting the transmit point to this image point will generate a intersection with the plane, which becomes the reflection point. As expected, considering more complex geometries, bigger scenarios, and higher reflection orders escalates computational efforts. The other \ac{RL} algorithm considers a large number of rays launched from a transmit point and shoot into the scenario. Upon intersecting with a plane, the hit point and the direction of the reflected ray are obtained and calculated. Unlike \ac{RT} methods, the receiver in \ac{RL} is modeled as a sphere with certain radius. If this reflected ray can hit the receiver sphere, it will be considered as a valid reflection path and this process is also named as \ac{SBR} method \cite{fuschini2015}. To speed up the calculations, investigations have been conducted on various acceleration strategies. Authors in \cite{8259292} employ the \ac{VPL} algorithm to divide a 3D simulation area into horizontal and vertical planes. Additionally, leveraging NVIDIA OptiX \cite{nvidia-optix} for parallel computing on GPUs, the program mentioned in \cite{felbecker2012electromagnetic} focus only on multiple reflections. The newly developed ray-launcher based on \cite{fugen2006capability} also utilizes NVIDIA OptiX and is used in this work, which has accelerated computation times significantly. For reference, a small urban area model, consisting of 2,557 polygons as in Fig. \ref{simulation_scene}, requires only about 4.5 seconds to compute for each frame, making it an extraordinary fast \ac{RL} tool. The simulation includes consideration of \ac{LOS} transmissions, up to second-order reflections, and diffractions.

In this paper, we will focus on the \ac{FR1} frequencies, and in particular the 3700-3800 MHz band. Authors in \cite{10278464} measured the \ac{SS-RSRP} within the RPTU Kaiserslautern \ac{PCN} and made their measurements dataset publicly available. The terrain, as shown in Fig. \ref{meas_scene}, is rather undulating, which can be observed from the yellow stair markers. Besides, it is lush with vegetation throughout the campus. This paper does not concentrate on the accuracy or proximity of our simulation results to the reported measurements. Instead, it illustrates how detailed channel information can be extremely beneficial in complex environments, presents methods for handling simulation data, and outlines approaches to address the issue of vegetation.

The rest of this paper is organized as follows. It starts with an overview of related works in Section \ref{related_works}. The simulation and measurement environment and parameters are briefly described in Section \ref{simulation_and_measurements}. Section \ref{results} introduces the results from our simulation and compares it to the reported measurements. A concise discussion of the findings follows in Section \ref{discussion}. Finally, conclusions are provided in Section \ref{conclusions}.

\section{Related Works} \label{related_works} 
In the spectrum allocated for \ac{5G} planning \cite{3gpp_38_104}, the 3300-3800 MHz band is particularly noteworthy within the sub-6G range because it has been widely adopted around the globe. On the one hand, numerous studies have been published that address measurement campaigns in urban outdoor environments operating at \ac{5G} \ac{FR1}. In \cite{he2016}, \ac{5G} channel measurements were conducted on a university campus in China at 3.5 GHz with a bandwidth of 30 MHz and the typical channel parameters were derived. Large-scale fading at same center frequency in an urban-macro scenario were measured and analyzed in \cite{zhang2019}. Whereas in \cite{adegoke2021}, small-scale fading effects were also included and the channel quality was investigated in terms of \ac{EVM}. In fact, for greater representatives and to derive wave propagation channels, studies mentioned above conducted measurements in flat terrains and open areas unlike the scenario represented by the measurements in \cite{10278464} that we base our analysis on.

On the other hand, using simulation techniques to gain insights of the wireless channel propagation and \ac{EMF} exposure patterns has also been addressed by several studies. In \cite{aguirre2014}, an estimation of the \ac{EMF} exposure levels for the indoor scenario of a commercial passenger airplane was achieved using a 3D \ac{RL}-based algorithm. The impact of reduced power levels due to \ac{EMF} exposure limits for \ac{5G} indoor scenarios was studied with the help of a \ac{RT} approach in \cite{wigren2015}. Authors in \cite{celaya_echarri2021} provided a comprehensive assessment of \ac{EMF} exposure in shopping malls, representing a high node density scenario in an indoor environment, by means of measurements as well as 3D \ac{RL}-based simulations. 

\section{Simulation and Measurements Description} \label{simulation_and_measurements}
\begin{figure*}[!t]
\centering
\subfloat[]{\includegraphics[width=3in]{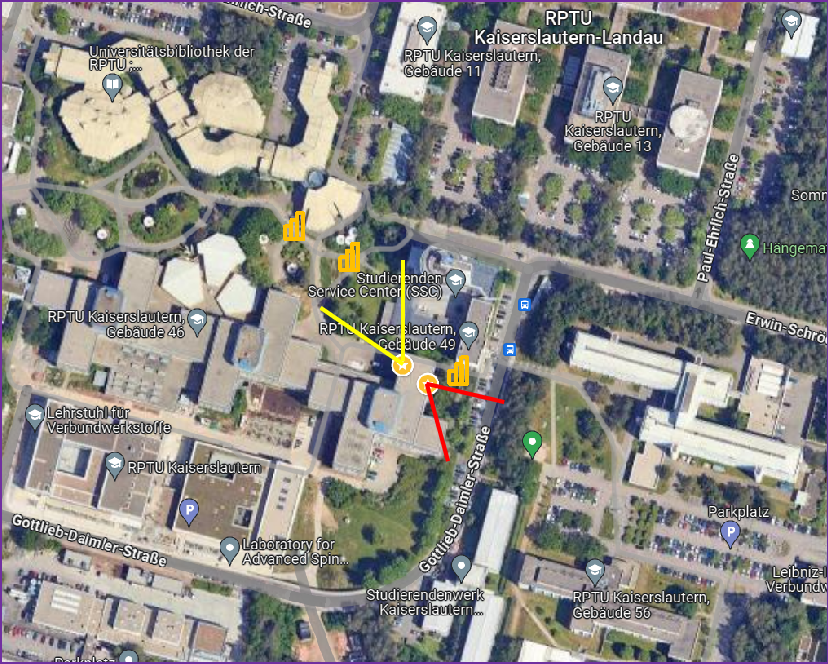}%
\label{meas_scene}}
\hfil
\subfloat[]{\includegraphics[width=3in]{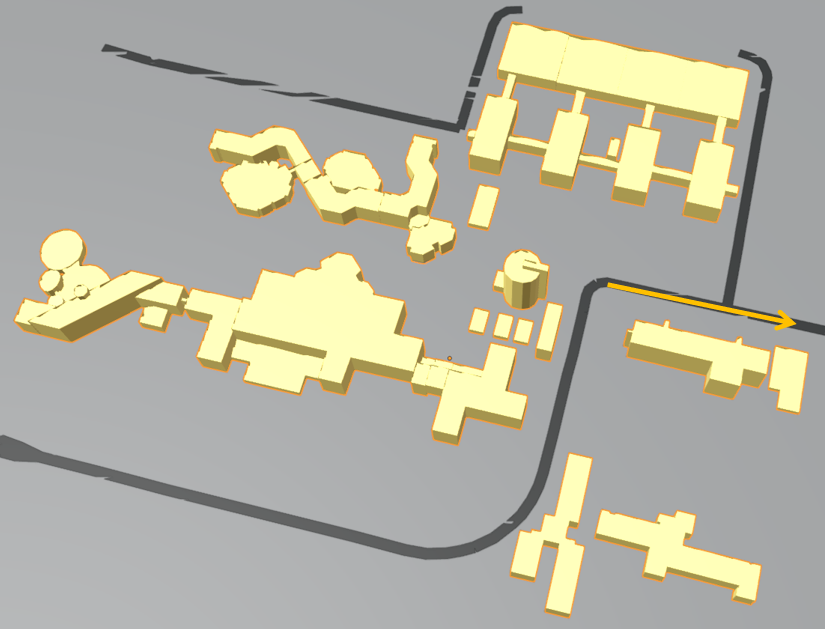}%
\label{simulation_scene}}
\caption{Environment characteristics of RPTU Kaiserslautern campus. (a) Measurement scenario from Google Maps. (b) Simulation scenario from Blender.}
\label{whole_scene}
\end{figure*}
In \cite{10278464}, three outdoor \acp{RRH} in total are deployed and their respective \ac{SS-RSRP} is measured. Among them, the \ac{BS} with \ac{PCI} 54 is selected for comparison because it is located in the central area of the campus as seen in Fig. \ref{meas_scene}. Each \ac{RRH} comprises two transmitters located in close proximity, each transmitter is equipped with a 2-port sector antenna that supports 2x2 \ac{MIMO} and operates in the 3.7-3.8 GHz frequency range. The azimuth and elevation \ac{HPBW} of these antennas are detailed in Table \ref{tab:parameter}, thereby covering the two sectors depicted in yellow and red as shown in Fig. \ref{meas_scene}. 
This surrounding is characterized by the most complex building structures, varied topological slopes, and an abundance of trees in its vicinity. The \ac{UE} ID is privately assigned within the \ac{PCN}. The \ac{UE} can navigate along various paths around the campus. Simulations are conducted based on the same scenario, with building geometry imported from \ac{OSM} as depicted in Fig. \ref{simulation_scene}. In this scenario, the existence of vegetation is initially not considered and the simulation parameters are based on those listed in Table \ref{tab:parameter}. Given the relatively narrow bandwidth, it is assumed that the channel is non-selective within this frequency range, and thus, only 3.75 GHz is simulated as the center frequency.

\begin{table}
\caption{Applied simulation environment parameters based on the measurement scenario according to \cite{10278464}}
\label{tab:parameter}
\centering
\begin{tabular}{|c|c|} 
\hline
Description & Value \\
\hline
Carrier frequency & 3.75 GHz\\
\hline
TX antenna position & 49°25'25.5"N 7°45'14.7"E \\
& and 49°25'25.2"N 7°45'15.4"E\\
\hline
TX antenna height & 22 m \\
\hline
TX output power & 20 W \\
\hline
TX antenna gain & 12.5 dBi \\
\hline
TX HPBW, & azimuth: $65^{\circ}$, elevation: $22^{\circ}$ \\
orientation (from north to east) and tilt & $330^{\circ}$, $10^{\circ}$ and $124^{\circ}$, $1^{\circ}$ \\
\hline
RX antenna height & 20 cm \\
\hline
RX antenna gain & 4 dBi \\ 
\hline 
RX antenna pattern & Omni-directional \\ 
and polarization &  vertical\\
\hline
\end{tabular}
\end{table}

In a \ac{RL} simulation, each ray undergoes specific attenuation, phase shift, and delay as it travels from the transmitter to the receiver, as a result of various propagation phenomena. The channel frequency response is calculated according to Equation (\ref{channel_response}), in which $n = 1,...N$ represents the different multi-paths. $G_{T}$, $G_{R}$ are the transmitter and receiver antenna gains, respectively. The complex directional pattern of the transmitter and receiver antennas are $C_{T}$ and $C_{R}$, respectively. Whereas, $\Omega_{T,n}$ and $\Omega_{R,n}$ represent the simulated \ac{AoD} and \ac{AoA} of a given path $n$. The full polarimetric transmission matrix is given by $T_{n}$ and the time delay is $\tau_{n}$. $c_{0}$ is the speed of light and $f_{c}$ is the center frequency. 
\begin{equation}
\label{channel_response}
\begin{aligned}
H(f) &= \sqrt{{(\frac{c_{0}}{4\pi f_{c}})}^2 G_R G_t}\cdot \sum_{n=1}^{N}C_{R}(\Omega_{R,n}) \cdot T_{n}C_{T}(\Omega_{T,n})e^{-j2\pi f \tau_{n}} \\
      &= \sum_{n=1}^{N} A_{n}e^{-j2\pi f \tau_{n}}.
\end{aligned}
\end{equation}
Where $A_{n}$ represents the complex amplitude of the $n^{th}$ multi-path. The final received power can be determined by coherently summing up all the path contributions based on Equation (\ref{channel_response}).

\section{Measurements and simulation results} \label{results}
In this section, the results of the \ac{RL} simulation and the reported measurements, where the \ac{UE} follows the same route indicated by the orange arrow in Fig. \ref{simulation_scene}, are presented. We will sequentially demonstrate how to deal with this complex scenario and take into account more and more levels of detail into our simulation.

\subsection{Case I: Simulation without trees}
\begin{figure}[!h]
\centering
\includegraphics[width=3.3in]{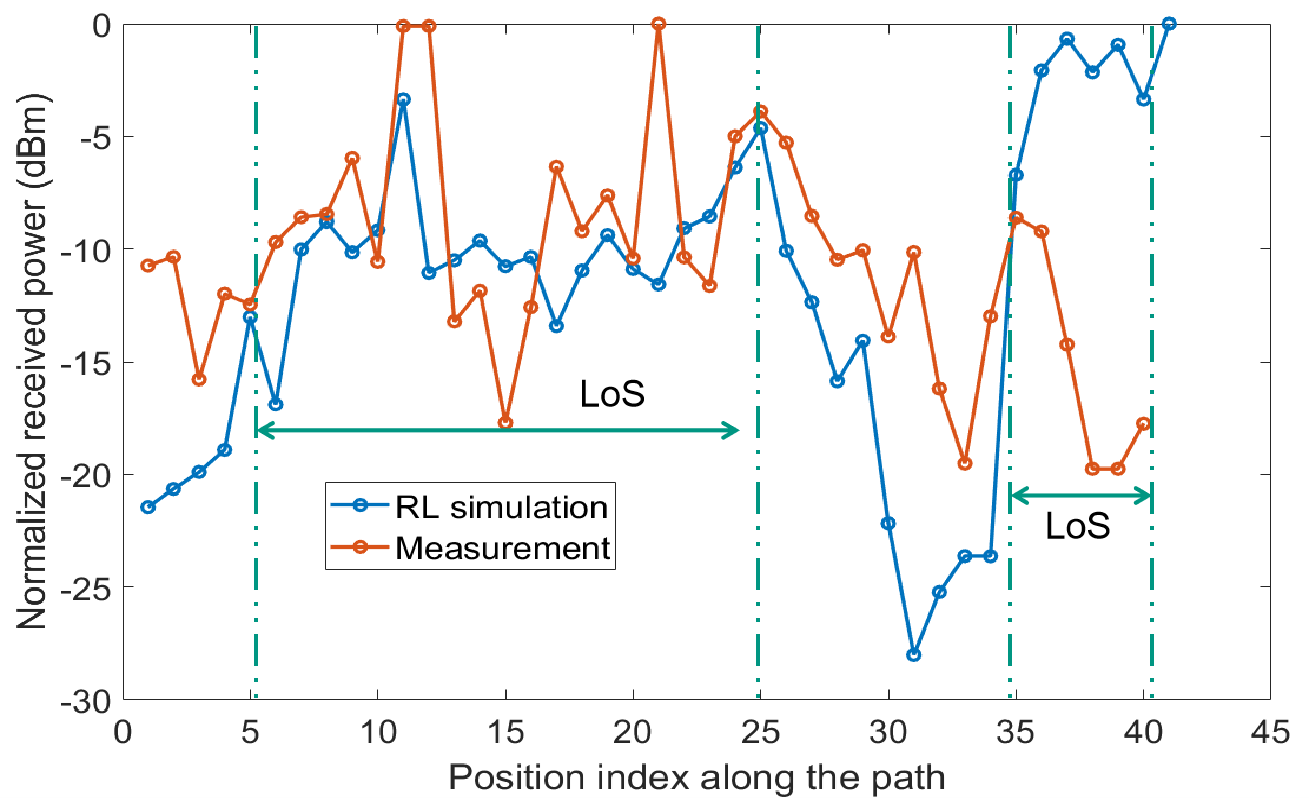}
\caption{Normalized received power from measurements and RL simulation}
\label{normlaized_no_tree}
\end{figure}
Fig. \ref{normlaized_no_tree} displays the normalized received power, with the maximum simulated received power being in -41 dBm and the maximum measured received power being -82 dBm. It is evident that the simulated received power is considerably higher than the measured values, attributed to the absence of occlusion from tall trees in this simulation scenario, thereby allowing for \ac{LOS} paths in between. However, this figure still demonstrates the alignment of the trend between the simulated and measured results.
This implies that the simulation results can be readily applied to optimize a \ac{BS}'s communication link to a \ac{UE} in urban environments. As the \ac{UE} travels through the city, the \ac{BS} can dynamically adjust its output power in response to the simulation outcomes. Moreover, this difference in the received power between our simulation and reported measurements could also vary according to other factors such as different weather conditions. However, it can be effectively calibrated by another neighboring \ac{BS}. 

Despite the absence of trees in this scenario, the simulation results also provide valuable insights into \ac{EMF} exposure issues. As suggested by the \ac{ICNIRP} in \cite{international1998guidelines}, it is imperative that the EMF exposure (electric field strength, magnetic field strength, power density or \ac{SAR}) limit must not be exceeded under any circumstances, necessitating the consideration of worst-case scenarios. Considering the significant seasonal foliage density variations, with trees being lush in summer and barren in winter, and given the ease of trees felling and planting, excluding the attenuation effect of trees would represent a worst-case scenario in assessing \ac{EMF} exposure risks.

\subsection{Case II: Simulation with trees} \label{case2_with_trees}
As illustrated by the actual scenario in Fig. \ref{meas_scene}, trees in the vicinity of the measurement route are exceptionally dense and form a continuous cover. Therefore, to align the simulation more closely with the real-world environment, the inclusion of dense trees is necessary. These trees have been incorporated into the simulation environment, represented by the green cuboids in Fig. \ref{simulation_scene_with_tree}. The \ac{UE} path is still represented by the orange arrow.
\begin{figure}[!t]
\centering
\includegraphics[width=3.5in]{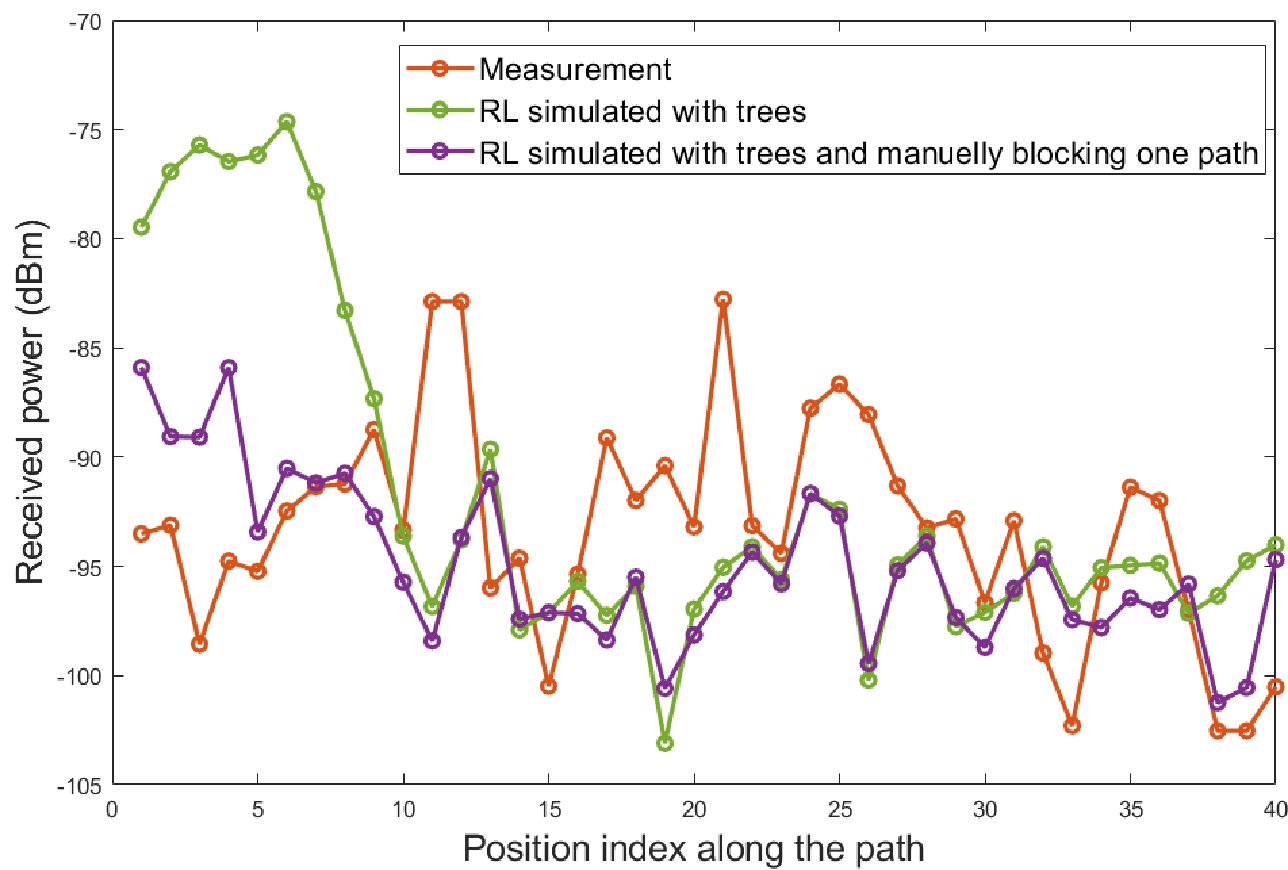}
\caption{Comparison of the measured and simulated received power with additional details (trees and blockers)}
\label{with_blocking_comparison}
\end{figure}
Due to their density, we assume that these trees do not allow any propagation paths and are treated as simple blockers. A comparison between the simulation results based on this environment and the actual reported measurements, as seen by the green and orange lines in Fig. \ref{with_blocking_comparison} respectively, reveals that the simulated received power is getting closer with the measured values in contrast to Fig. \ref{normlaized_no_tree}. However, it can be observed that at the beginning of the path, the simulated received power still significantly exceeds the measured value. This discrepancy is attributed to the presence of a strong first-order diffraction path, as depicted by the red line with arrow in Fig. \ref{simulation_scene_with_tree}. 
Towards the end of the \ac{UE} route along the street, the simulated received power slightly underestimates the measured values. This is due to the tall trees acting as blockers on both sides, which over-suppress the signal paths. In reality, certain paths can still penetrate between the trunk or through the foliage and reach the \ac{UE}.

\subsection{Case III: Manual blocking}
\begin{figure*}[!htbp]
\centering
\subfloat[]{\includegraphics[width=3in]{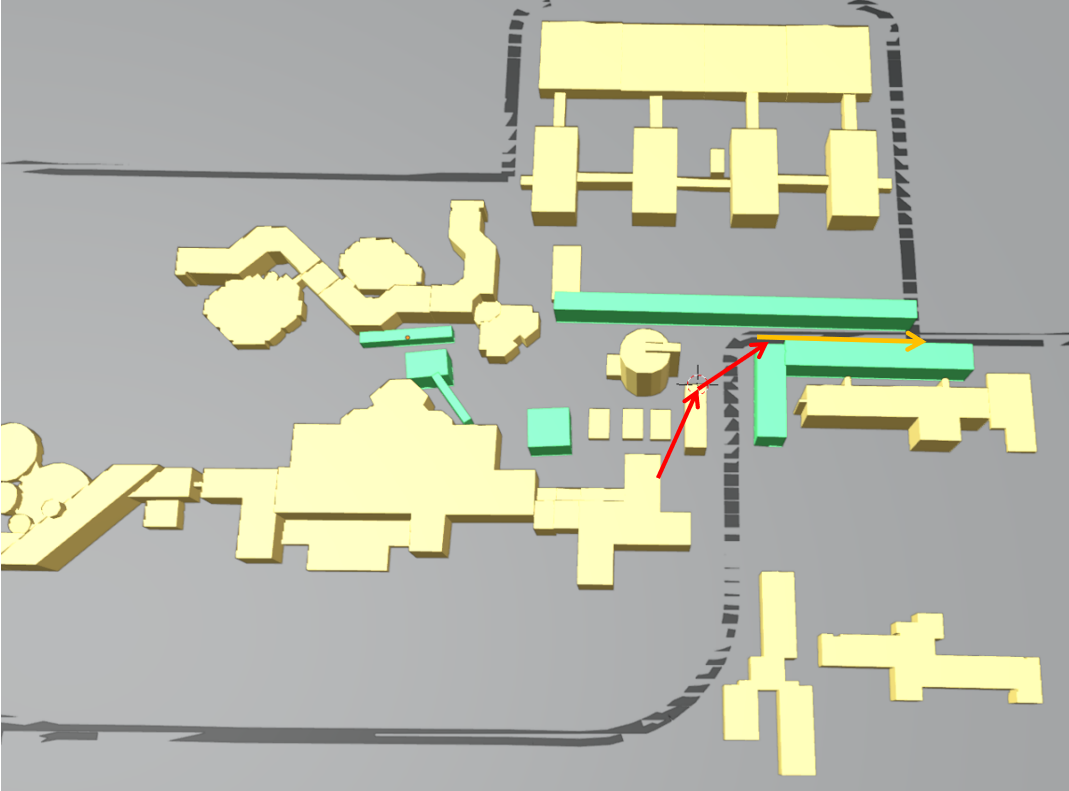}%
\label{simulation_scene_with_tree}}
\hfil
\subfloat[]{\includegraphics[width=3.5in]{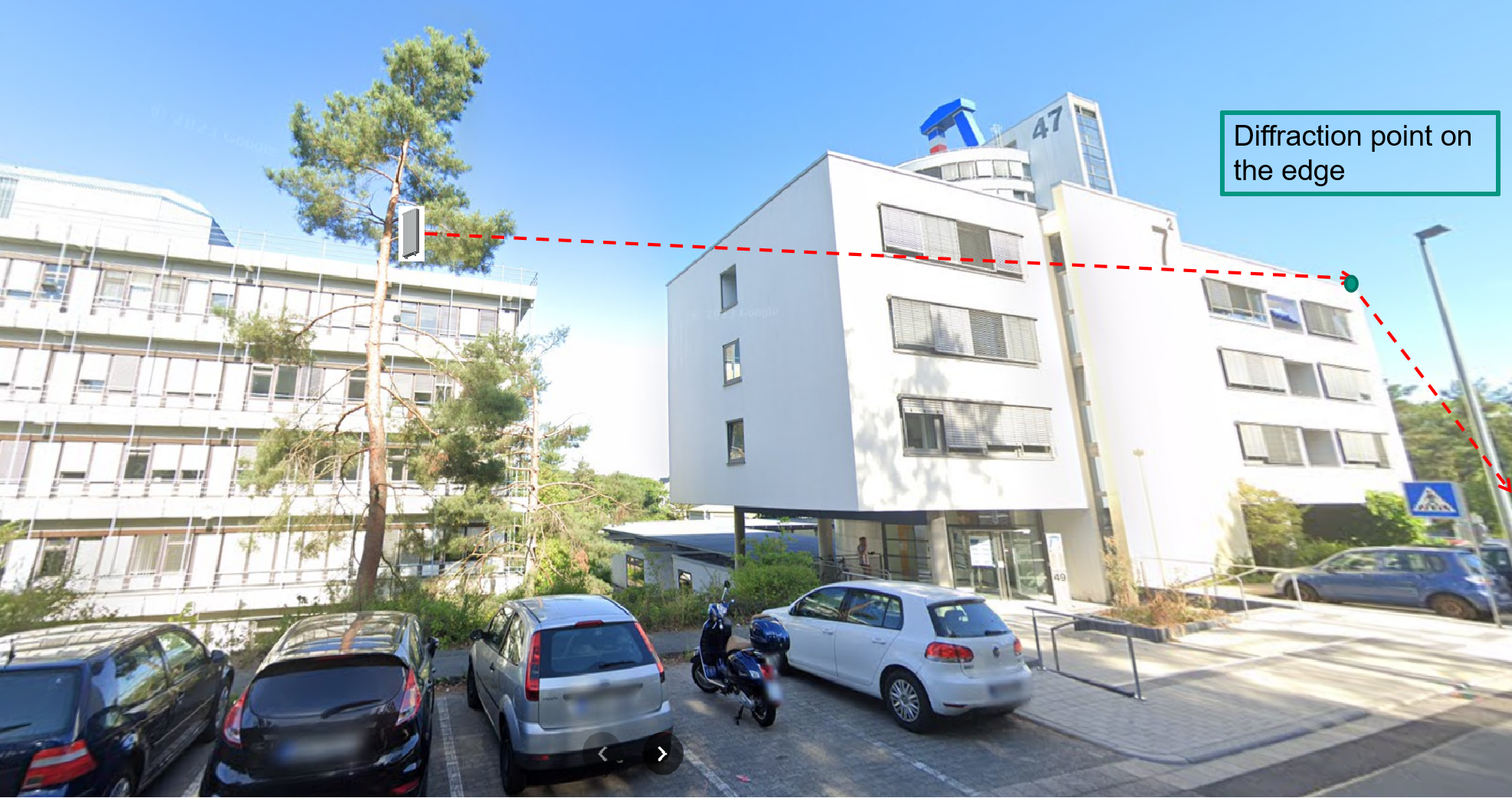}%
\label{diffraction_reality}}
\caption{(a) Simulation scenario from Blender with trees. (b) Reality of the buildings from Google Maps. The dashed red line represents the strong diffraction path, which is similarly depicted as the solid red arrow line in (a).}
\label{second part}
\end{figure*}

By integrating real-world environmental factors, the simulation results can be further refined. As illustrated in Fig. \ref{diffraction_reality}, the actual environment features a building on the left with a \ac{RRH} transmitter on the rooftop. Adjacent to it on the right is a building associated with the diffraction path mentioned in the subsection \ref{case2_with_trees}, which is depicted by a red dashed line in Fig. \ref{diffraction_reality}. However, it can be spotted that the building on the left is constructed on a lower level, indicating the terrain in this area is not flat. Nevertheless, acquiring precise information about terrain undulations is always challenging. Normally, assistance from the Land Bureau or other governmental agencies would be required to obtain accurate terrain models. Even in the absence of terrain information, this issue can be adequately addressed as follows. The imported building models in Fig. \ref{simulation_scene_with_tree} are based on the same ground level, and only the height information of the buildings can be obtained. Nevertheless, utilizing \ac{RL} simulation, specific interaction points can be obtained. Given the variance in ground elevation, the altitude of the \ac{RRH} transmitter atop the left building is not higher than the right building, therefore, the diffraction path can be excluded reasonably. By manually removing this path, the simulation result, as shown by the violet curve in Fig. \ref{with_blocking_comparison}, align more closely with the measured results, particularly evident in the initial positions of route.

\section{Discussion} \label{discussion}
Multiple positions for a \ac{UE} traveling in the campus area have been measured and simulated. Measurement results over the territory are illustrated in Fig. \ref{with_blocking_comparison}a, whereas simulated results for the entire area, without any trees, are depicted in Fig. \ref{with_blocking_comparison}b. 
\begin{figure}[!t]
\centering
\includegraphics[width=3.5in]{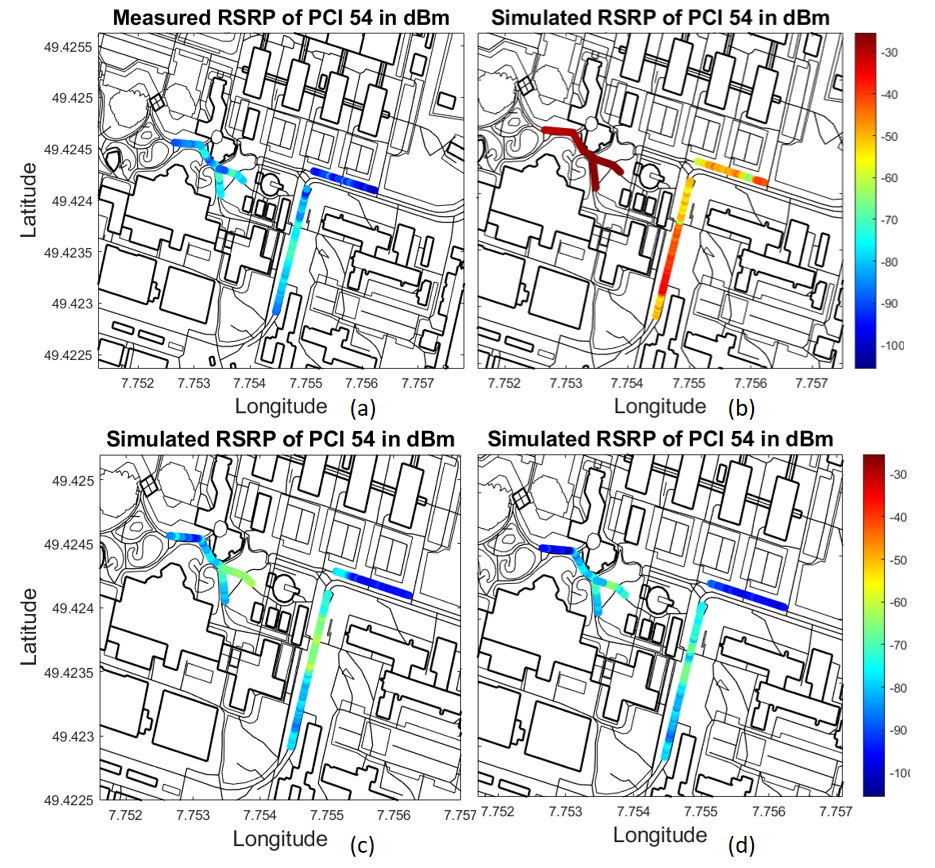}
\caption{(a) Measured received power. (b) Simulated received power without any trees. (c) Simulated received power with trees. (d) Simulated received power with trees and manually blocking.}
\label{with_blocking_comparison}
\end{figure}
As expected, all simulated results in Fig. \ref{with_blocking_comparison}b exceed the measured values, particularly for the positions in the upper left, which are close to the \ac{RRH} and possess a strong \ac{LOS} path. Across the 168 positions evaluated, the \ac{RMSE} between simulated and measured results is calculated to be 43.4 dB, significantly higher than the measured values but the overall trend still aligns as shown in Section III.A. Following the incorporation of trees as described in Section III.B, the simulated results, presented in Fig. \ref{with_blocking_comparison}c, exhibit a reduction in the \ac{RMSE} to 9.9 dB, closely approximating the measured values. Further adjustments to block certain paths considering reality, as outlined in Section III.C, result in the simulated outcomes depicted in Fig. \ref{with_blocking_comparison}d, achieving an \ac{RMSE} of only 7.5 dB when compared to the measurements.

\section{Conclusions} \label{conclusions}
In this paper, the benefit of utilizing the \ac{RL} method for simulating the received power in the sub-6 GHz band has been demonstrated. A complex scenario within the university campus of RPTU Kaiserslautern, characterized by uneven terrain and an abundance of trees, was intentionally chosen. Simulations were initially conducted for scenarios without considering any trees. Although the simulated received power was higher than that measured, the former followed the overall trend of the latter. Subsequently, trees were incorporated as blockers, bringing the simulation results closer to the measured ones, albeit with some discrepancies. By leveraging detailed \ac{RL} outputs to identify interaction points and manually removing paths that could not exist in reality, the simulation results were significantly aligned with the measured data. Further simulations across various positions achieved an overall \ac{RMSE} of 7.5 dB, which closely approximates the measurements. This step-by-step improvement highlights the necessity of incorporating a higher level of detail within the \ac{RL} simulation scenario to achieve simulation results that closely reflect the actual measurements.

\section*{Acknowledgments}
The authors acknowledge the financial support by the Federal Ministry of Education and Research of Germany (BMBF) in the project “Open6GHub” (grant numbers: 16KISK010 and 16KISK004). Thanks also go to RPTU Kaiserslautern for giving the full measurement details and all the other co-authors who has helped to review the paper. 

%

\bibliography{mybibliography.bib} 
\bibliographystyle{ieeetr} 

\end{document}

%% file: acronyms.tex
\DeclareAcronym{3G}{short = 3G , long = third generation}
\DeclareAcronym{3GPP}{short = 3GPP , long = Third Generation Partnership Project}
\DeclareAcronym{4G}{short = 4G , long = fourth generation}
\DeclareAcronym{5G}{short = 5G , long = fifth generation}
\DeclareAcronym{6G}{short = 6G , long = sixth generation}

\DeclareAcronym{AoA}{short = AoA , long = angle of arrival}
\DeclareAcronym{AoD}{short = AoD , long = angle of departure}
\DeclareAcronym{ASiR}{short = ASiR , long = AirScale indoor radiohead}

\DeclareAcronym{BBU}{short = BBU ,  long = baseband unit}
\DeclareAcronym{BS}{short = BS ,  long = base station}

\DeclareAcronym{CN}{short = CN , long = core network}

\DeclareAcronym{EIRP}{short = EIRP , long = effective isotropic radiated power}
\DeclareAcronym{eMBB}{short = eMBB , long = enhanced mobile broadband}
\DeclareAcronym{EMF}{short = EMF , long = electromagnetic field}
\DeclareAcronym{ESIM}{short = ESIM , long = earth stations in motion}
\DeclareAcronym{ETSI}{short = ETSI , long = European Telecommunications Standards Institute}
\DeclareAcronym{EU}{short = EU , long = European Union}
\DeclareAcronym{EVM}{short = EVM , long = error vector magnitude}

\DeclareAcronym{FCC}{short = FCC , long = Federal Communications Commission}
\DeclareAcronym{FDD}{short = FDD , long = frequency division duplex}
\DeclareAcronym{FR1}{short = FR1 , long = frequency range 1}
\DeclareAcronym{FR2}{short = FR2 , long = frequency range 2}
\DeclareAcronym{FSS}{short = FSS , long = fixed satellite service}

\DeclareAcronym{GEO}{short = GEO , long = geostationary earth orbit}
\DeclareAcronym{GO}{short = GO , long = geometrical optics}
\DeclareAcronym{HAPS}{short = HAPS , long = high altitude platform station}
\DeclareAcronym{HPBW}{short = HPBW , long = half-power beamwidth}

\DeclareAcronym{ICES}{short = ICES , long = International Committee on Electromagnetic Safety}
\DeclareAcronym{ICNIRP}{short = ICNIRP , long = International Commission on Non-Ionizing Radiation Protection}
\DeclareAcronym{IEC}{short = IEC , long = International Electrotechnical Commission}
\DeclareAcronym{IEEE}{short = IEEE , long = Institute of Electrical and Electronics Engineers}
\DeclareAcronym{IMT}{short = IMT , long = International Mobile Telecommunications}
\DeclareAcronym{IoT}{short = IoT , long = Internet of Things}
\DeclareAcronym{ITU}{short = ITU , long = International Telecommunication Union}

\DeclareAcronym{LEO}{short = LEO , long = low earth orbit}
\DeclareAcronym{LOS}{short = LOS , long = line of sight}
\DeclareAcronym{LTE}{short = LTE , long = long term evolution}

\DeclareAcronym{MEO}{short = MEO , long = medium earth orbit}
\DeclareAcronym{mMIMO}{short = mMIMO , long = massive multiple input multiple output}
\DeclareAcronym{mMTC}{short = mMTC , long = massive machine type communication}
\DeclareAcronym{mmWave}{short = mmWave , long = millimeter wave}
\DeclareAcronym{MIMO}{short = MIMO , long = massive multiple input multiple output}
\DeclareAcronym{mRRH}{short = mRRH , long = micro-RRH}
\DeclareAcronym{MSS}{short = MSS , long = mobile satellite service}

\DeclareAcronym{NATO}{short = NATO , long = North Atlantic Treaty Organization}
\DeclareAcronym{NCRP}{short = NCRP , long = National Council on Radiation Protection and Measurements}
\DeclareAcronym{NDAC}{short = NDAC , long = Nokia Digital Automation Cloud}
\DeclareAcronym{NG-RAN}{short = NG-RAN , long = next generation radio access network}
\DeclareAcronym{NGSO}{short = NGSO , long = non-geostationary satellite orbit}
\DeclareAcronym{NR}{short = NR , long = new radio}
\DeclareAcronym{NSA}{short = NSA , long = non-stand-alone}
\DeclareAcronym{NTN}{short = NTN , long = non-terrestrial network}

\DeclareAcronym{OSM}{short = OSM , long = OpenStreetMap}

\DeclareAcronym{PBCH}{short = PBCH , long = physical broadcast channel}
\DeclareAcronym{PCI}{short = PCI , long = physical cell ID}
\DeclareAcronym{PCN}{short = PCN , long = private campus network}
\DeclareAcronym{PDSCH}{short = PDSCH , long = physical downlink shared channel}
\DeclareAcronym{pRRH}{short = pRRH , long = pico-RRH}

\DeclareAcronym{QoE}{short = QoE , long = quality of experience }
\DeclareAcronym{QoS}{short = QoS , long = quality of service }

\DeclareAcronym{RAN}{short = RAN , long = radio access network}
\DeclareAcronym{RAT}{short = RAT , long = radio access technology}
\DeclareAcronym{RF}{short = RF , long = radio frequency}
\DeclareAcronym{RRH}{short = RRH , long = remote radio heads}
\DeclareAcronym{RSS}{short = RSS , long = root sum square}
\DeclareAcronym{RL}{short = RL , long = ray-launching}
\DeclareAcronym{RT}{short = RT , long = ray-tracing}
\DeclareAcronym{RMSE}{short = RMSE , long = root mean square error}

\DeclareAcronym{SA}{short = SA , long = stand-alone}
\DeclareAcronym{SAN}{short = SAN ,  long = satellite access node}
\DeclareAcronym{SAR}{short = SAR , long = specific absorption rate}
\DeclareAcronym{SBR}{short = SBR , long = shooting and bouncing rays}
\DeclareAcronym{SSB}{short = SSB , long = synchronization signal block}
\DeclareAcronym{SS-RSRP}{short = SS-RSRP , long = synchronization signal - reference signal received power}

\DeclareAcronym{TDD}{short = TDD , long = time division duplex}
\DeclareAcronym{THz}{short = THz , long = terahertz}
\DeclareAcronym{TN}{short = TN , long = terrestrial network}
\DeclareAcronym{TUK}{short = TUK, long = Technische Universit\"at Kaiserslautern}

\DeclareAcronym{UAS}{short = UAS , long = unmanned aerial system }
\DeclareAcronym{UE}{short = UE , long = user equipment}

\DeclareAcronym{VSAT}{short = VSAT , long = very small aperture terminal }
\DeclareAcronym{VPL}{short = VPL , long = vertical-plane-launch }

\DeclareAcronym{WHO}{short = WHO , long = World Health Organization}

